\documentclass[fleqn,usenatbib]{mnras}

\usepackage{newtxtext,newtxmath}

\usepackage[T1]{fontenc}

\DeclareRobustCommand{\VAN}[3]{#2}
\let\VANthebibliography\thebibliography
\def\thebibliography{\DeclareRobustCommand{\VAN}[3]{##3}\VANthebibliography}

\usepackage{graphicx}	
\usepackage{amsmath}	

\title{Is quasar variability regulated by the close environment of accretion?}

\author[L. Wu et al.]{
Liang Wu$^{1,2}$\thanks{E-mail: wul@mail.ustc.edu.cn},
Jun-Xian Wang$^{1,2}$\thanks{E-mail: jxw@ustc.edu.cn}, 
Wen-Ke Ren$^{1,2}$,
Wen-Yong Kang$^{1,2}$
\\
$^{1}$CAS Key Laboratory for Research in Galaxies and Cosmology, Department of Astronomy, University of Science and Technology of
China, Hefei, Anhui 230026, \\China\\
$^{2}$School of Astronomy and Space Science, University of Science and Technology of China, Hefei 230026, China
}

\pubyear{2024}

\begin{document}
\label{firstpage}
\pagerange{\pageref{firstpage}--\pageref{lastpage}}
\maketitle

\begin{abstract}
UV/optical variability in quasars is a well-observed phenomenon, yet its primeval origins remain unclear. This study investigates whether the accretion disk turbulence, which is responsible for UV/optical variability, is influenced by the close environment of the accretion by analyzing the correlation between variability and infrared emission for two luminous SDSS quasar samples. The first sample includes light curves from SDSS, Pan-STARRS, and ZTF $g$ band photometry, while the second sample utilizes SDSS Stripe 82 $g$ band light curves. We explore the correlation between the $g$ band excess variance ($\sigma_{rms}$) and the wavelength-dependent infrared covering factor ($L_{\rm IR}(\lambda)/L_{\rm bol}$), controlling for the effects of redshift, luminosity, and black hole mass. An anti-correlation between two variables is observed in both samples, which is strongest at wavelengths of 2-3$\micron$ but gradually weakens towards longer wavelength. 
This suggests the equatorial dusty torus (which dominates near-infrared emission) plays a significant role in influencing the UV/optical variability, while the cooler polar dust (which contributes significantly to mid-infrared emission) does not. 
The findings indicate that quasar variability may be connected to the physical conditions within the dusty torus which feeds the accretion, and support the notion that the close environment of the accretion plays an important role in regulating the accretion disk turbulence.
\end{abstract}

\begin{keywords}
galaxies: active —- quasars: general
\end{keywords}



\section{Introduction}

Variability is a critical topic in the studies of active galactic nuclei (AGNs)
and quasars, which exhibit intense aperiodic variations across the entire
spectrum, from radio waves to X-rays and gamma-rays 
\citep[e.g.][]{ulrich_variability_1997}. 
The UV/optical emission from the accretion disk surrounding the supermassive black hole is the primary continuum radiation of AGNs and quasars. Studying its variation, which could be due to thermal fluctuations driven by the turbulent magnetic field in the disk \citep[e.g.][]{Kelly2009}, thus is essential to probe the yet unclear nature of the accretion process. 

Attributing the UV/optical variability to magnetic turbulence in the disk, investigating the correlation between variability and other processes could yield interesting clues about what conditions drive stronger turbulence or the consequences of the turbulence.
Notably, recent works have revealed intrinsic correlations between UV/optical variability amplitude and X-ray loudness \citep{kang_var_x-ray_link_2018}, and emission line strength \citep{kang_var_line_2021}. While these studies indicate both the heating of the X-ray corona and the launch of emission line clouds in quasars could be physically associated with the magnetic turbulence in the accretion disk, they also demonstrate the feasibility and efficiency to probe the consequences of disk turbulence through studying the correlation between variability and other processes utilizing large time-domain data sets.

However the primeval origin of the variability is yet unclear. Could the observed variability be solely be attributed to local stochastic magnetic processes in the disk, or somehow related to the environment which feeds the accretion as the fluctuations at larger scales may slowly propagate into the inner disk? 
Note the inward propagation of disk fluctuations appears able to explain the observed low frequency X-ray lags in 
X-ray binaries and AGNs \citep[e.g.][]{Kotov2001,Arevalo2006}. This propagation has also been possibly revealed in the optical multi-band light curves in individual AGN \citep{Neustadt2024}. If 
the environment feeding the accretion does play a role in regulating the inner disk fluctuations, we would expect a correlation between variability and the environment in large samples of AGNs. 

Meanwhile, numerous studies have revealed anti-correlation between UV/optical 
variability and bolometric luminosity or Eddington ratio
\citep[e.g.][]{
wold_variability_mass_2007,
wilhite_edd_ratio_variability_2008,
macleod_variability_2010,
zuo_variability_2012,
meusinger_variability_accretion_rate_2013, Kozlowski_variability_sf_2016, 
sun_evolution_variability_2018, 
kang_var_x-ray_link_2018,Petrecca2024}, though the underlying physics is yet unclear, which also suggests a connection between variability and the feeding to the accretion \citep[e.g.][]{wilhite_edd_ratio_variability_2008}. 

An essential component of the AGN unification model is the toroidal dusty structure in the equatorial plane, known as the torus
\citep[e.g.][]{antonucci_unified_1993, urry_unified_1995}. This structure not only
absorbs radiation from the nuclear region and re-emits it in the infrared
band, but also serves as a local gas reservoir to fuel the black hole accretion. 
It is thus intriguing to explore the correlation between UV/optical variability and torus emission to probe the physical connection between the turbulence of the accretion disk and its close dust environment. 
The relative strength of torus emission is generally quantified with $L_{IR}/L_{bol}$, where $L_{IR}$ measures the dust emission and $L_{bol}$ the bolometric luminosity. However, recent studies have revealed the ubiquitous existence of polar dust in AGNs and quasars \citep{lopez_polar_2016, asmus_polar_2016, asmus_polar_2019, wu_ensemble_2023}. Unlike the equatorial torus, the polar dust is located at larger distance and along the polar direction, thus is unlikely to feed the central accretion.
Therefore, when quantifying the equatorial torus emission with $L_{IR}/L_{bol}$, the contribution from polar dust need to be carefully subtracted. 
Though the polar dust is located on larger scales (compared with the inner torus), likely associated with disk winds or dusty NLR \citep{netzer_nlr_dust_1993, honig_polar_2019, Stalevski2019}, spatially resolving and subtracting the polar dust emission is extremely challenging. Fortunately, as demonstrated by \cite{wu_ensemble_2023}, because the polar dust is cooler, its contribution to near infrared emission is minimal while it may dominate the infrared emission at 10 $\micron$. Therefore near infrared emission could be taken as the best proxy for the equatorial torus emission.

In this work, we for the first time study the intrinsic correlation between UV/optical variability and infrared emission, using large samples of SDSS quasars with time-domain observations.   The paper is organized as follows: In \S2, we describe the samples, 
the measurements of variability amplitude $\sigma_{\rm rms}$ and the relative strength of infrared emission $L_{\rm IR}(\lambda)/L_{\rm bol}$. 
In \S3, we present the correlation analyses and the scientific results. 
Discussion and conclusions are given in \S4.

\section{Sample and measurements}

\begin{figure}
    \centering
    \includegraphics[width=0.49\textwidth]{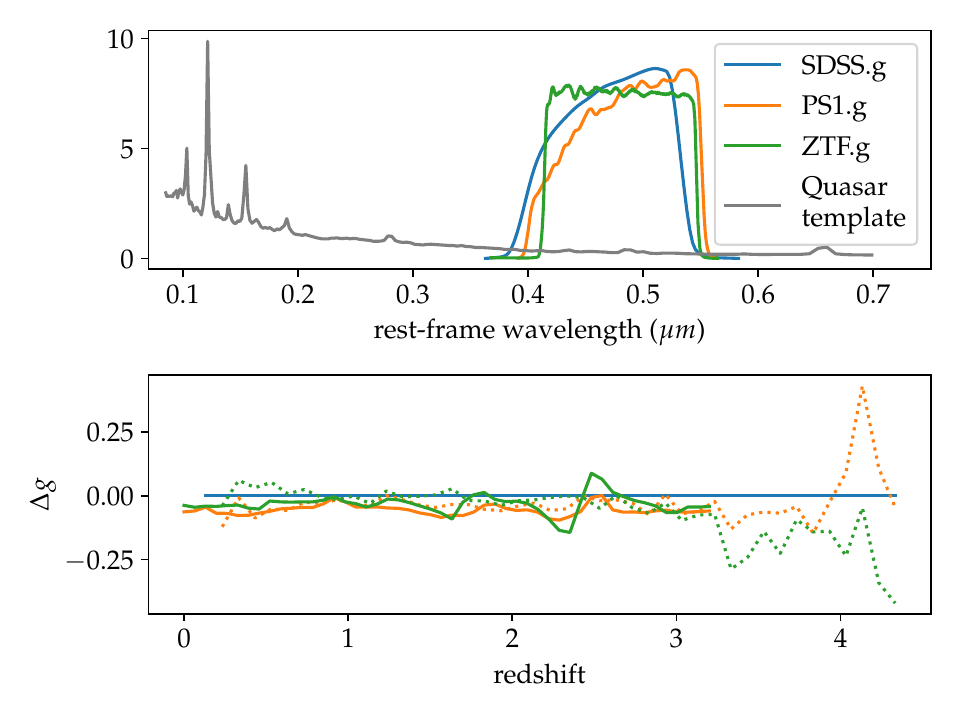}
    \caption{
        Upper panel: Normalised $g$ band transmission curves of SDSS, PS1 and ZTF. A quasar template from \protect\cite{shang_new_2011} is shown in gray for comparison.
        Lower panel: Photometric $g$ band difference caused by filter. The solid lines represent $g$ band photometric difference (compared to SDSS $g$) computed through convolving the quasar template with the transmission curve of each filter, while the dotted lines represent directly measured mean magnitude difference for quasars in our full DR7 sample obtained in each redshift bin of 0.1. 
    }
    \label{fig:filter}
\end{figure}

\begin{figure}
    \centering
    \includegraphics[width=0.5\textwidth]{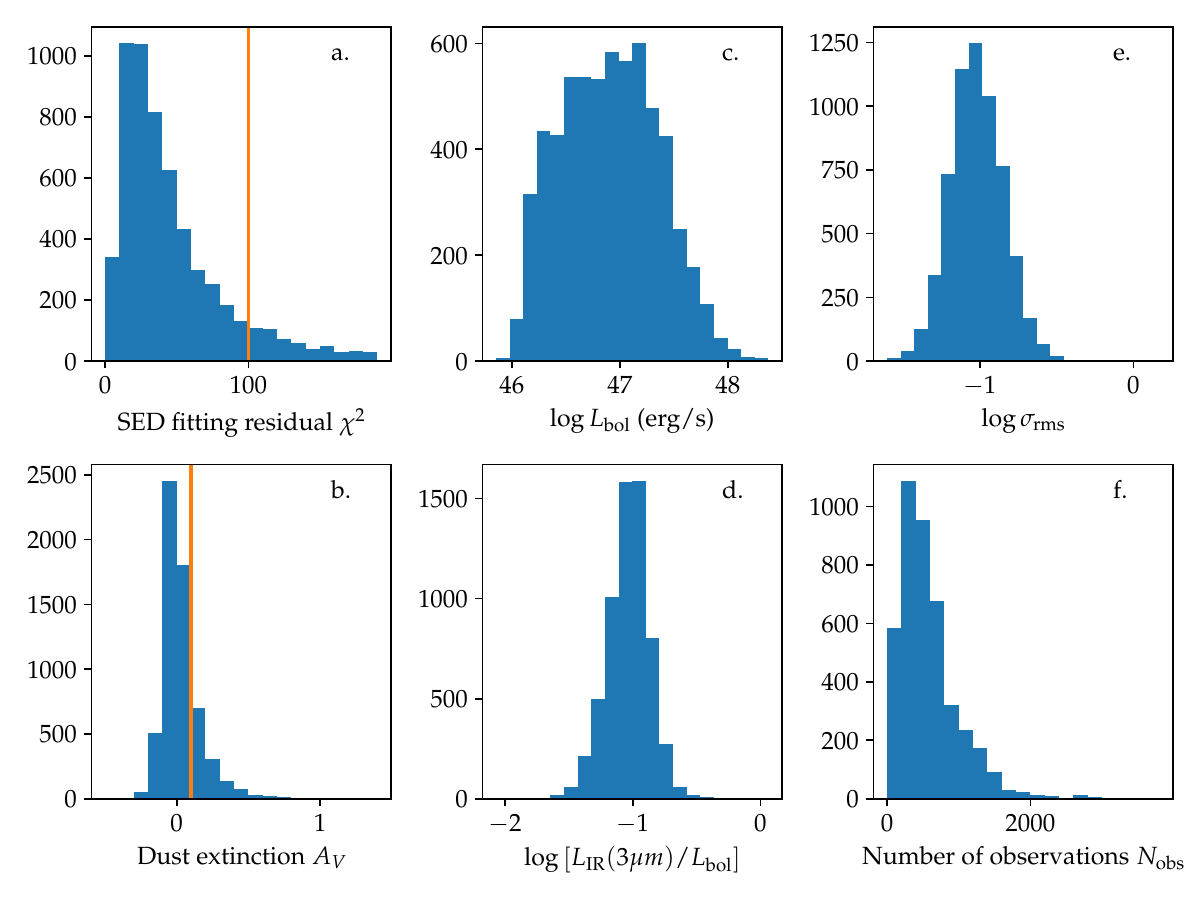}
    \caption{
        Measurements of the DR7 sample. (a) SED fitting residual of hot+cold dust model. The orange vertical line is located at residual equals 100, and sources to its right are dropped. (b) Dust extinction $A_V$ obtained by SED fitting. The orange vertical line is located at $A_V$ equals -0.1, and sources to its right are dropped.  Note around half of our quasars exhibit weak negative extinction because they have UV/optical SED bluer than the disk template \citep{shang_new_2011} we adopted. (c) Bolometric luminosity $L_{\rm bol}$ derived from interpolated $L_{5100}$. (d) 3$\mu m$ covering factor derived from interpolated bolometric luminosity and 3$\mu m$ luminosity. (e) Excess variance. (f) Total number of observations from SDSS, PS1 and ZTF.
    }
    \label{fig:measurements}
\end{figure}

\begin{figure}
    \centering
    \includegraphics[width=0.5\textwidth]{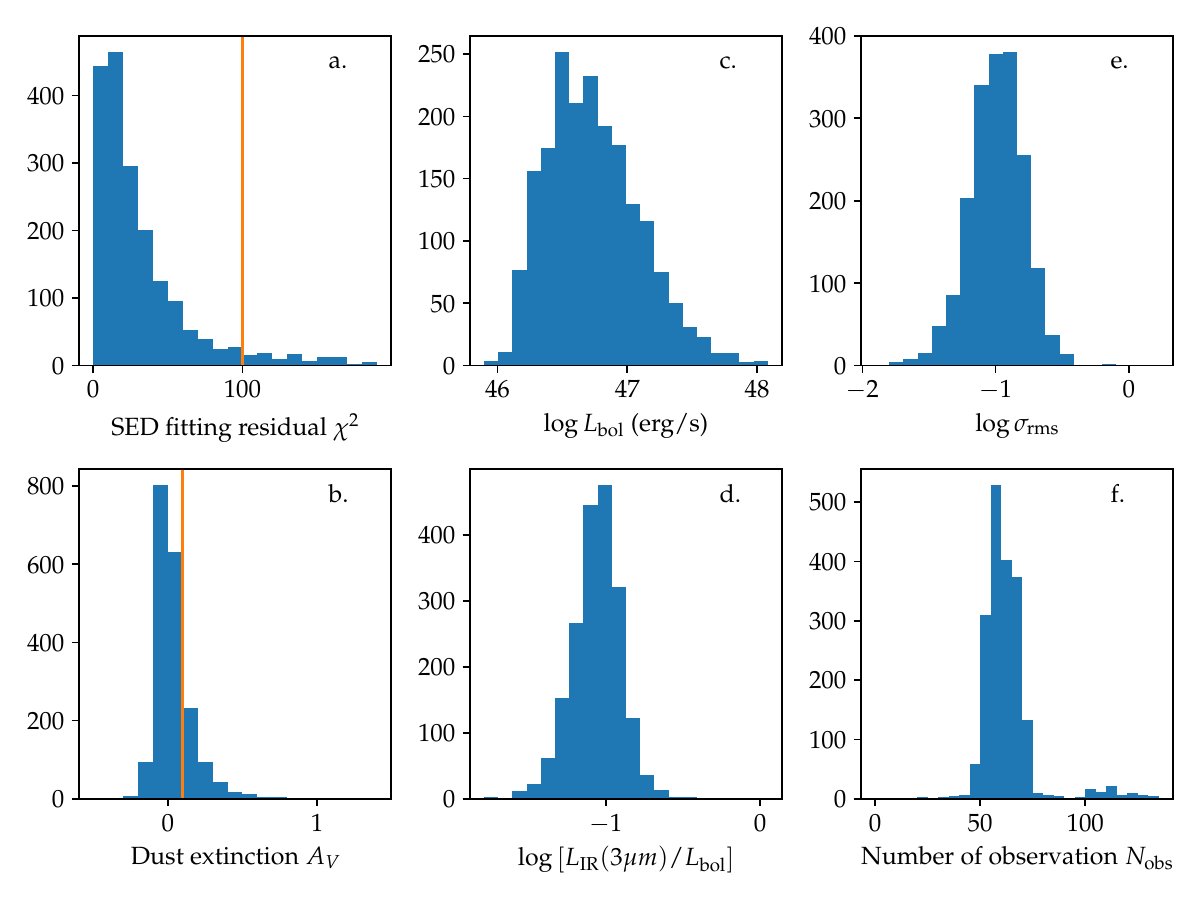}
    \caption{
        Same as Fig. \ref{fig:measurements} but for the Stripe 82 sample. 
    }
    \label{fig:measurements_s82}
\end{figure}

\begin{figure}
    \centering
    \hspace{-8mm}
    \includegraphics[width=0.52\textwidth]{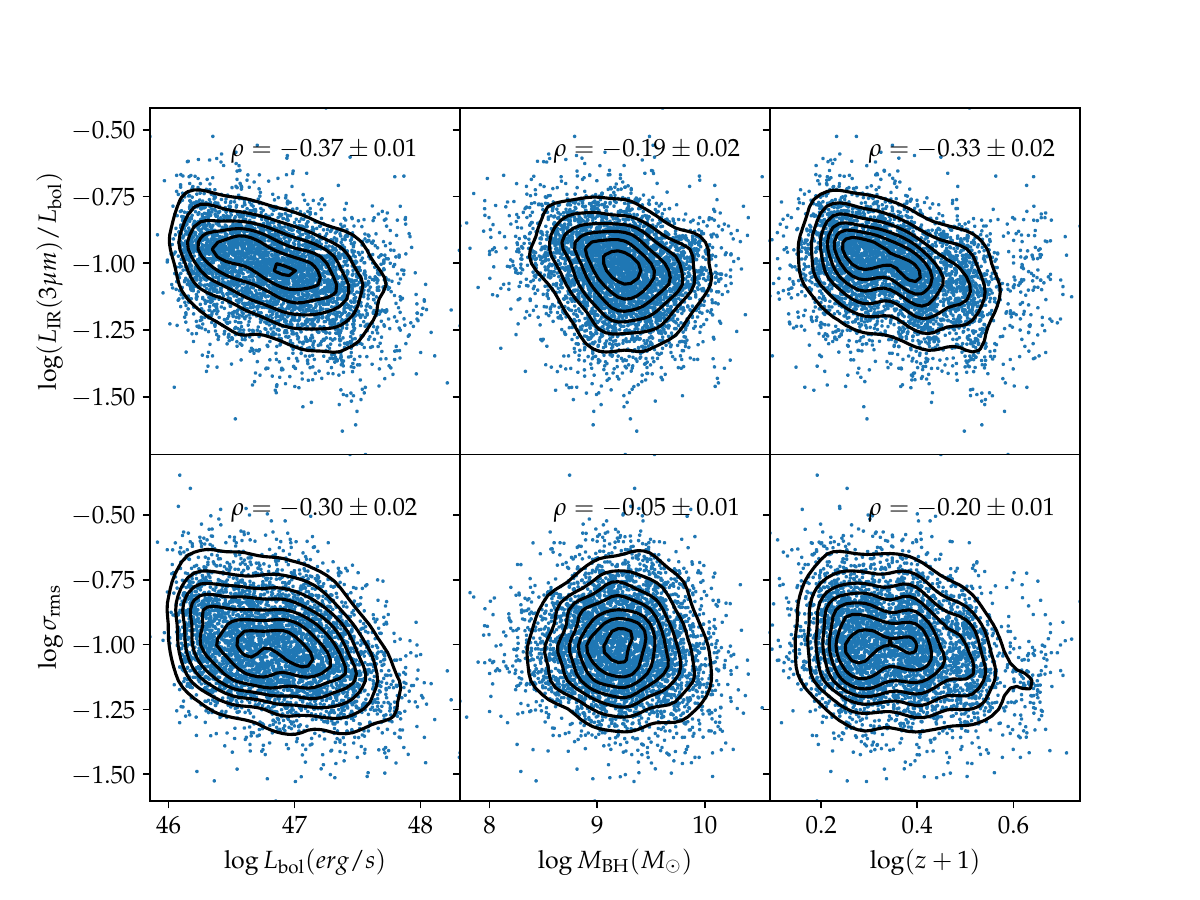}
    \caption{ $g$-band
        $\sigma_{\rm rms}$ (upper) and $L_{\rm IR}(3\mu m)/L_{\rm bol}$ (lower) versus $L_{\rm bol}$, $M_{\rm BH}$ and redshift, respectively, for the full DR7 sample. The Spearman correlation coefficients $\rho$ between the quantities and their errors calculated through bootstrap are labelled for each panel. 
    }
    \label{fig:corrs}
\end{figure}

\subsection{The Data and Samples}

Following \cite{wu_ensemble_2023}, this work focuses on quasars with bolometric luminosity above $10^{46}$ erg/s to avoid significant host galaxy contamination in the photometry. We begin with the SDSS DR7\footnote{Using the latest SDSS data releases only slightly increases the sample size of luminous quasars (e.g., see \citealt{wu_ensemble_2023}), but involves considerable additional effort. Meanwhile, multi-band photometry, uniform SDSS spectral fitting results, and optical to infrared SED fitting are directly available for SDSS DR7 quasars from \cite{shen_catalog_2011} and \cite{wu_ensemble_2023}, respectively.} quasar catalog \citep{schneider_sdss_dr7_2010} and collect their multi-band photometry, including data from SDSS \citep{fukugita_sdss_1996}, 2MASS \citep{carpenter_2mass_2001}, and WISE \citep{wright_wise_2010}. 

Additionally, we utilize SDSS spectral fitting results (luminosity and supermassive black hole mass) from \cite{shen_catalog_2011}.

\subsubsection{The SDSS Stripe 82 Sample}

We consider two samples of SDSS quasars with time-domain photometry. For SDSS quasars in Stripe 82 (a 290 deg$^2$ equatorial field of the sky; \citealt{Sesar2007}), which have been scanned over 60 times in SDSS $ugriz$ bands, we utilize the re-calibrated 10-year SDSS light curves presented by \cite{Macleod2012}. We use SDSS $g$-band photometry for the following reasons:

1) Photometric Uncertainties: The $g$ and $r$ bands have the smallest photometric uncertainties among the five SDSS bands ($ugriz$), making them more reliable for variability studies \citep{SunYH2014}.

2) Intrinsic Variability: Quasars exhibit stronger intrinsic variability at shorter wavelengths \citep{meusinger_variability_2011, zuo_variability_2012, Petrecca2024}, making the $g$ band more suitable for this study compared to the $r$ band.

Note that the $g$ band of quasars at different redshifts corresponds to different rest-frame wavelengths. When the redshift range of the sample is large, direct comparison is not feasible. However, this effect can be controlled through partial correlation analysis, as detailed in \S 3.

We exclude quasars without WISE photometry, resulting in a sample of 1939 luminous quasars with bolometric luminosity above $10^{46}$ erg/s, as reported by \cite{shen_catalog_2011}. Among these, 308 quasars also have 2MASS photometry. Although 2MASS photometry is not essential for this study, it is included when available for subsequent spectral energy distribution (SED) fitting.

\subsubsection{The full DR7 sample}

Though the light curves of Stripe 82 sample are of high quality ($\sim$ 10 years long from a single instrument), the limited sky coverage reduce its sample size significantly. Thus we decide to construct another sample from the full SDSS DR7 sample with light curves obtained from archived photometric data from SDSS, Pan-STARRS, and ZTF. 
For the full quasar sample of DR7, the sample size increases significantly that a large sample of 8120 sources is obtained after excluding quasars with bolometric luminosity below $10^{46}$ erg/s and requiring observations from both WISE and 2MASS. 
However, this catalog provides only single-epoch photometry (i.e., the primary SDSS magnitude) for each source.
To construct $g$ band light curves for each quasar, we 
gather all available archival photometric observations from SDSS\footnote{\url{https://skyserver.sdss.org/casjobs/}}, 
Pan-STARRS1\footnote{\url{https://mastweb.stsci.edu/ps1casjobs/home.aspx}} 
(PS1, \citealt{chambers_ps1_2016}) and ZTF \citep{bellm_ztf_2019}\footnote{\url{https://irsa.ipac.caltech.edu/Missions/ztf.html}}
databases. 
This comprehensive approach allows us to analyze the variability characteristics of a larger sample of quasars over extended periods.

Archival SDSS photometric observations are
obtain from the DR14 database, where the primary photometric observations 
were obtained with SDSS-I/II, lasting from 2000 to 2007 (covering 11,663 deg$^2$)
, and SDSS-III (before 2009) on $\sim$3000 deg$^2$ of new sky area. Following
\cite{ren_evq_2022}, we gather all available SDSS $g$ band photometry using a matching radius of 1$^{\prime\prime}$.  
In accordance with recommendations for utilizing photometric processing flags from SDSS\footnote{\url{https://live-sdss4org-dr16.pantheonsite.io/algorithms/photo_flags_recommend/}}, 
we retain detections with mode = 1 or 2, "clean" flags, and point spread
function (PSF) magnitude error $<$ 0.2 mag in the $g$ band.
Approximately 51\% quasars have multi-epoch SDSS photometry, with 268 of them observed over more than 5 epochs.
Note 
Stripe 82 detections are not part of the DR14 database but are in a separate database. To avoid selection bias, we do not supplement them to the DR7 sample.

Following \cite{ren_evq_2022}, we also collect the $g$ band photometry from 
the PS1 3$\pi$ survey, which conducted up to four exposures in each band per year from 2009 to 2013. The matching radius is also set to 
1$^{\prime\prime}$ and matched PS1 detections are also filtered according to
their photometric info flags. We rule out PS1 detections with
PSF magnitude error > 0.2 mag or flagged as follows:

\begin{enumerate}
    \item Peak lands on diffraction spike, ghost or glint;
    \item Poor moments for small radius, try large radius or could not measure the moments;
    \item Source fit failed or succeeds but with low signal-to-noise ratio (S/N), high Chi square or too large for PSF;
    \item Source model peak is above saturation;
    \item Size could not be determined;
    \item Source has crNsigma above limit;
    \item Source is thought to be a defect;
    \item Failed to get good estimate of object's PSF;
    \item Detection is astrometry outlier.
\end{enumerate}

Eventually, 78\% quasars (6,325 quasars) in our DR7 sample have PS1 detection, with a  mean of approximately 10.3 epochs in the $g$ band.

The Zwicky Transient Facility (ZTF) is a new time-domain survey that has 
achieved first light in November of 2017. 
As the successor to PTF and iPTF, ZTF maintains PTF’s image quality over a much wider field and provides more than an order of magnitude increase in survey speed. 
The photometric data were taken by 16 6k $\times$ 6k e2V CCD with a 47 square 
degree field of view mounted on the Samuel Oschin 48-inch Schmidt telescope \citep{Harrington1952}.
ZTF observations are also obtain by setting a matching radius 
of 1$^{\prime \prime}$ and mask of 65535 \footnote{Only detections with no flag 
would be retained, see 
\url{https://irsa.ipac.caltech.edu/data/ZTF/docs/ztf_explanatory_supplement.pdf} 
for flag definitions.}. 
Thanks to the large sky coverage of ZTF, 97\% quasars (6152 quasars) are detected
in ZTF from 2018 to 2023 with a mean number of observations around 540. The distribution of the total number of $g$ band observations from SDSS, PS1 and ZTF is shown in Fig. \ref{fig:measurements}.

Finally, we must consider the differences in the filter transmission between SDSS, PS1 and ZTF, before constructing the
$g$ band light curves. 
We first consider the zero point differences between telescopes. According to the SVO Filter Profile Service\footnote{\url{http://svo2.cab.inta-csic.es/theory/fps/index.php}} \citep{rodrigo_svo_2012}, Vega zero points of each telescope are 5.45476e-9 (erg/cm2/s/A, SDSS), 5.05397e-9 (erg/cm2/s/A, PS1) and 5.2673e-9 (erg/cm2/s/A, ZTF), so the magnitude difference caused by the zero point difference is 0.08 between SDSS and PS1,   and 0.04 between SDSS and ZTF. We first apply a correction to PS1 and ZTF data to eliminate this offset.
In Fig. \ref{fig:filter}, we plot the $g$ band full transmission curves of each telescope and a quasar template from \cite{shang_new_2011} for reference. The magnitude difference caused by different filter transmission is redshift dependent.  Convolving the filter transmission with
the quasar template, we obtain the expected magnitude difference
($\Delta g=g_*-g_{\rm SDSS}$) as shown in the lower panel of Fig. \ref{fig:filter}, which appears well consistent with the directly measured mean magnitude difference for our quasars in the full DR7 sample obtained in each redshift bin of 0.1. 
As the quasar template extends down to
only 0.09$\micron$ in the rest frame, resulting in a cut at $z\sim3.2$ in the output magnitude difference,  
we choose to apply a correction to PS1 and ZTF photometry data using the directly measured mean magnitude difference.

\subsection{The Excess Variance $\sigma_{\rm rms}$}

In recent years, the damped random walk (DRW) process has been commonly adopted to model quasar light curves, with the DRW parameters (the damping timescale $\tau$ and SF$_\infty$) derived to represent quasar variability \citep[e.g.,][]{Kelly2009, zu_drw_2013, Kozlowski_variability_sf_2016, suberlak_drw_2021}. However, the measurements of these two parameters can be severely biased if the light curves are poorly sampled or if the baseline is not sufficiently long compared to $\tau$ \citep{Kozlowski2017, Hu2024ApJ}. This issue is particularly challenging for the luminous quasars used in this work, as their characteristic variation timescale $\tau$ is expected to be even longer, and the sampling for a substantial number of quasars, especially for the DR7 sample, is rather poor (see Fig. \ref{fig:measurements}). Moreover, it is still debated whether the DRW model sufficiently describes quasar variability \citep[e.g.,][]{Guo2017}.

Therefore, following \cite{kang_var_x-ray_link_2018,kang_var_line_2021}, we adopt the excess variance $\sigma_{\rm rms}$ \citep[e.g.,][]{vaughan_variability_2003} to quantify variability:
$$
\sigma_{\rm rms}^2=
\frac{1}{N-1} \sum_i (X_i - \bar{X})^2 - 
\frac{1}{N} \sum_i \sigma_i^2
$$
where $N$ is the total number of observations, $X_i$ is the $i$th observed magnitude, $\bar{X}$ is the average magnitude, and $\sigma_i$ is the photometric uncertainty of the $i$th observation. The statistical uncertainty of the excess variance (in the case of weak intrinsic variation) is given by:
$$
\mathrm{err}(\sigma^2_{\rm rms}) = 
\sqrt{\frac{2}{N}} \times \frac{1}{N} \sum_i \sigma_i^2
$$

In this work, we use $g$-band magnitudes measured by SDSS and Pan-STARRS to calculate $\sigma_{\rm rms}$ and represent quasar variability. Although $\sigma_{\rm rms}$ is robust for sources with a small number of observations, an adequate number of observations is necessary for $\sigma_{\rm rms}$ to reliably represent intrinsic variability. Therefore, we exclude sources with fewer than 20 observations in the $g$-band light curve.

Due to the statistical uncertainty mentioned above, there are 16 sources in the DR7 sample and 27 sources in the Stripe 82 sample with a negative $\sigma^2_{\rm rms}$, which are consequently excluded. Since these sources constitute only a minor fraction of the samples, their exclusion does not alter the main results presented in this work.

\subsection{The $L_{\rm IR}(\lambda)/L_{\rm bol}$}

SED fitting is a common approach to derive $L_{\rm IR}(\lambda)/L_{\rm bol}$, an indicator of the relative strength of infrared emission \citep[e.g.,][]{roseboom_ir-derived_2013,ma_covering_2013,toba_how_2021,wu_ensemble_2023}. However, as shown in \cite{wu_ensemble_2023}, this approach can be dependent on the chosen model or template. Furthermore, it remains challenging to properly decompose various dust components (e.g., separating polar dust from torus emission). 
In this work, following \cite{wu_ensemble_2023}, we simply interpolate the photometric data to obtain $L_{\rm IR}(\lambda)$ and use the interpolated $L_{5100}$ along with a bolometric correction factor of 9.26 \citep{shen_catalog_2011} to derive the bolometric luminosity $L_{\rm bol}$.

Nevertheless, we still perform SED fitting on the sample to exclude quasars with abnormal photometry or significant reddening, which could severely bias the interpolated $L_{5100}$. Following \cite{roseboom_ir-derived_2013} and \cite{wu_ensemble_2023}, we employ a model known as the hot+cold dust model for SED fitting. This model comprises three components: a disc, hot dust, and cold dust, with an SMC-like extinction applied (see \citealt{wu_ensemble_2023} for more details). For most sources, this model yields reasonable fits (see Fig. 1 \& 2 of \citealt{wu_ensemble_2023}). The distribution of the resulting $\chi^2$ and the extinction $A_V$ are shown in panels a \& b of Fig. \ref{fig:measurements} and \ref{fig:measurements_s82}.

It's worth noting that $\chi^2$ values (with a median of 31.9 and a mean of 103.0 for DR7 sample) are considerably larger than those reported by \cite{roseboom_ir-derived_2013}, likely because \cite{roseboom_ir-derived_2013} may have included additional photometric uncertainties in the SED data points (see also \citealt{wu_ensemble_2023}). A small fraction of sources are poorly fitted with $\chi^2 > 100$, primarily due to strong extinction that may not be well-fitted with a simple SMC extinction curve. For comparison, the mean $A_V$ of sources with SED fitting $\chi^2 > 100$ is 0.13, while for the total sample, it is 0.031.

Thus, we exclude sources with $\chi^2 > 100$ or $A_V > 0.1$ from the subsequent analysis, although including them would not alter the main results presented in this work. Ultimately, 4216 sources from the DR7 sample and 1427 sources from the Stripe 82 sample are retained.

\section{Correlation Analysis}

Correlation analysis is an efficient method for understanding the relationship between physical quantities. Among several correlation coefficients, we chose the Spearman correlation coefficient for this work because it is non-parametric and less sensitive to strong outliers.

It is well known that the UV/optical variability of quasars depends on rest-frame wavelength (being more variable at shorter wavelengths; e.g., \citealt{meusinger_variability_2011,zuo_variability_2012,Petrecca2024}) and timescale (being more variable at longer timescales; e.g., \citealt{deVries2005,Macleod2012,Guo2017,Petrecca2024}). Therefore, when performing correlation analysis between variation amplitude and other parameters, we must control for the effects of rest-frame wavelength and timescale. This can be achieved by controlling for the redshift parameter in partial correlation analysis, which eliminates the effects of time dilation and redshift-dependent rest-frame wavelength. At a given rest-frame wavelength, the characteristic variation timescale of quasars could depend on the supermassive black hole mass $M$ and bolometric luminosity $L_{\rm bol}$. Thus, controlling for $M$ and $L_{\rm bol}$ allows us to reveal the intrinsic correlation between variability and other quasar properties.

First, we examine the direct correlation between each of the two parameters of interest in this work (the 3 $\micron$ infrared covering factor $L_{\rm IR}(3\micron)/L_{\rm bol}$ and $g$-band excess variance $\sigma_{\rm rms}$) and several major quasar parameters, including bolometric luminosity $L_{\rm bol}$, black hole mass $M_{\rm BH}$, and redshift $z$. A $2 \times 3$ scatter plot with contours and Spearman correlation coefficients is shown in Fig. \ref{fig:corrs}. 

In our sample, $L_{\rm IR}(3\micron)/L_{\rm bol}$ shows a clear anti-correlation with both $L_{\rm bol}$ and $M_{\rm BH}$, consistent with previous studies \citep[e.g.,][]{ma_covering_2013, toba_how_2021, wu_ensemble_2023}. Additionally, $\sigma_{\rm rms}$ shows an anti-correlation with $L_{\rm bol}$ and a mild positive correlation with $M_{\rm BH}$. 
Regarding redshift, $L_{\rm IR}(3\micron)/L_{\rm bol}$ appears to evolve with redshift, but there is no obvious evolution of $\sigma_{\rm rms}$ with redshift. As both $L_{\rm IR}(3\micron)/L_{\rm bol}$ and $\sigma_{\rm rms}$ are correlated with fundamental quasar parameters, it is necessary to perform partial correlation analysis to avoid spurious correlation results.

\begin{figure}
    \centering
    \includegraphics[width=0.5\textwidth]{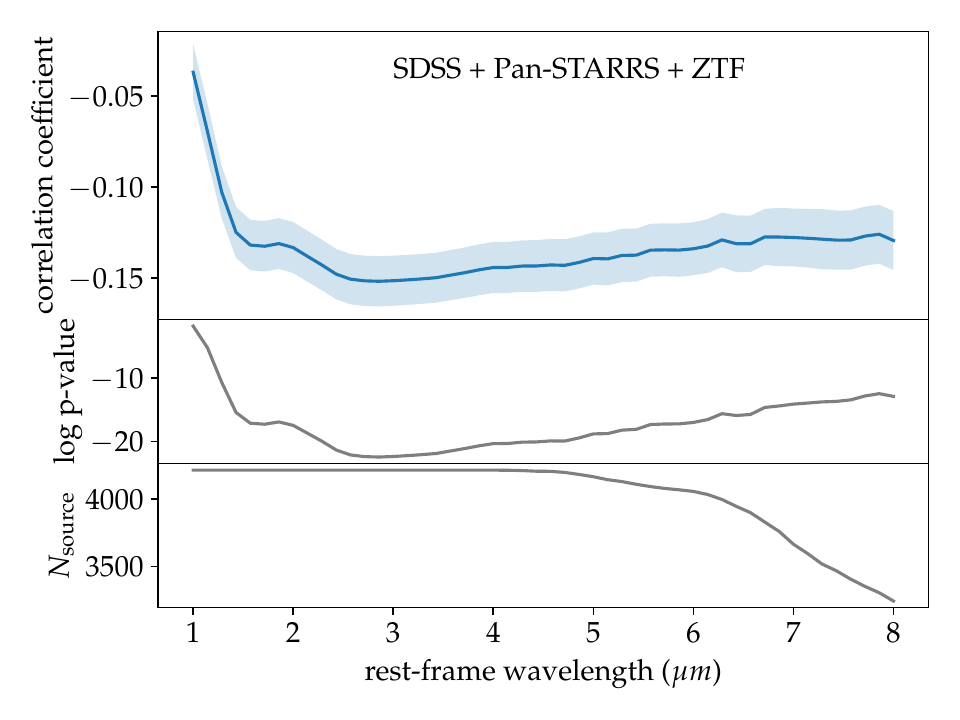}
    \caption{
        Upper: Partial correlation coefficient $\rho$ between $\sigma_{\rm rms}$ and $L_{\rm IR}(\lambda)/L_{\rm bol}$ as a function of $\lambda$ over rest-frame 1–8 $\mu m$. Middle: The p-value as a function of $\lambda$. Lower: Sample size as a function of $\lambda$. 
    }
    \label{fig:wav_corr}
\end{figure}

\begin{figure}
    \centering
    \includegraphics[width=0.5\textwidth]{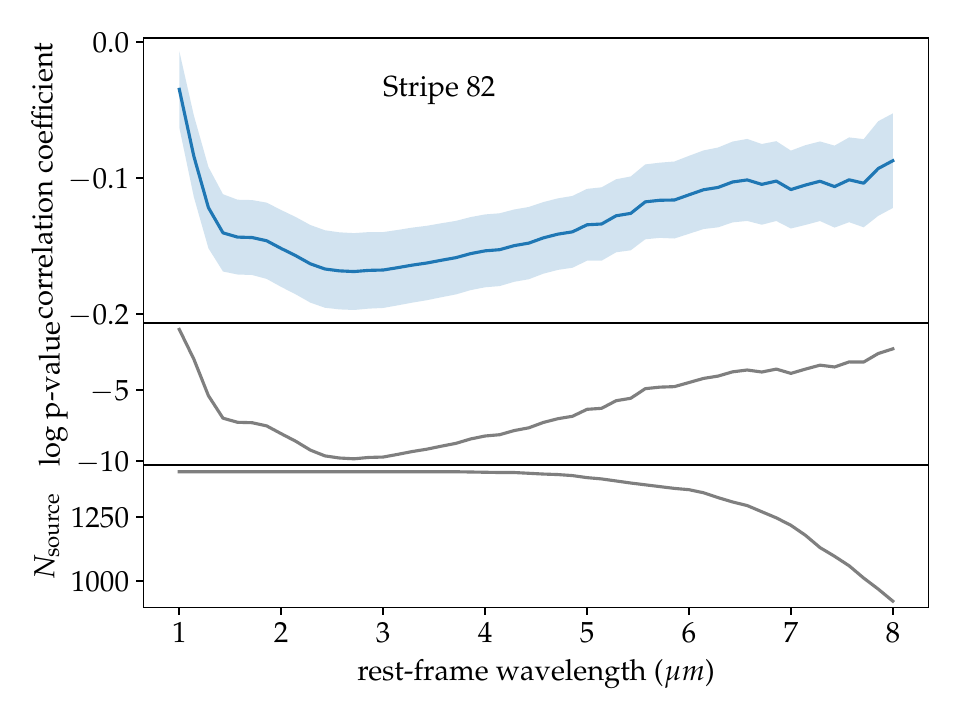}
    \caption{
        Same as Fig. \ref{fig:wav_corr} but for the Stripe 82 sample.
    }
    \label{fig:s82}
\end{figure}

We perform a partial Spearman correlation analysis between $L_{\rm IR}(\lambda)/L_{\rm bol}$ and $\sigma_{\rm rms}$ for wavelengths from 1 $\micron$ to 8 $\micron$, controlling for fundamental quasar parameters such as $L_{\rm bol}$, $M_{\rm BH}$, and redshift. The results for both the DR7 sample and the Stripe 82 sample are shown in Fig. \ref{fig:wav_corr} and Fig. \ref{fig:s82}, respectively. 

For both samples, the partial correlation coefficient decreases as the wavelength increases and reaches its minimum value of approximately -0.15 at around 3 $\micron$, and then slowly increases as the wavelength increases. Note that at $\lambda$ < 2-3$\micron$, the disc component in the SED is not negligible, and the shorter the wavelength, the more dominant the disc component becomes. Consequently, $L_{\rm IR}(\lambda)/L_{\rm bol}$ degenerates to a constant if $L_{\rm IR}(\lambda)$ is dominated by the disc component, leading to a lack of correlation with any quantity. This is why the partial correlation coefficient tends to zero at $\lambda$ < 2-3\micron\ (see also Fig. 3 of \citealt{wu_ensemble_2023}).

Note that some sources are discarded due to the lack of rest frame infrared wavelength coverage caused by high redshift, resulting in a decline in sample size at wavelengths greater than 4 $\micron$. The p-values, calculated based on the sample size and the correlation coefficient at each wavelength, are also shown in Fig. \ref{fig:wav_corr} and Fig. \ref{fig:s82} as a function of wavelength. 

Since the size of the DR7 sample (4216) is much larger than that of the Stripe 82 sample (1395), the anti-correlation for the DR7 sample is more significant (with p-value down to $1 \times 10^{-20}$) than those for the Stripe 82 sample (with p-value down to $1 \times 10^{-10}$).

Thanks to the large size of the DR7 sample, we are able to further divide it into different redshift bins. This allows us to confirm that our key results presented in Fig. \ref{fig:wav_corr} remain robust across a wide range of redshift (see Appendix \ref{AppendA} for details). Additionally, we divide the DR7 sample into various bins based on apparent magnitude, as the photometric errors could vary greatly with apparent magnitude. We find this step does not alter our key results either (see Appendix \ref{AppendB}). 

\section{Discussion}\label{sec:discussion}

In many aspects, the observed UV/optical variability of quasars could be well reproduced with the inhomogeneous disk model, which attributes the UV/optical variability to thermal (likely magnetic turbulence driven) fluctuations in the disk \citep{ dexter_disk_inhomogeneous_2011, cai_simulating_2016, cai_euclia_model_2018, Cai2020}. 
Particularly, the original inhomogeneous disk model of \cite{dexter_disk_inhomogeneous_2011} could naturally explain the discrepancy in the disk size between the thin disk model and that inferred from micro-lensing events \citep{dexter_disk_inhomogeneous_2011}, and the color variability \citep{Ruan2014}. More remarkably, the revised inhomogeneous disk models are able to explain the observed timescale-dependence of the color variability \citep{cai_simulating_2016}, and the observed inter-band lags and coordination without requiring light echoing \citep{cai_euclia_model_2018,Cai2020}. 

Within the scheme of inhomogeneous disk model, below we explore the physical nature of the anti-correlation between variability and infrared emission we discovered. \cite{cai_simulating_2016} predicted with the inhomogeneous disk model that quasars with stronger variability should have bluer SED. Though it is to be observationally verified, \cite{kang_var_x-ray_link_2018,kang_var_line_2021} found more variable quasars tend to have slightly stronger X-ray emission and UV/optical emission lines, suggesting more variable quasars may have harder SED. Harder SED could lead to relatively stronger infrared emission. However, this trend, if exists, contradicts the anti-correlation between the variability and infrared emission we detected in this work. 

\begin{figure}
    \centering
    \includegraphics[width=0.5\textwidth]{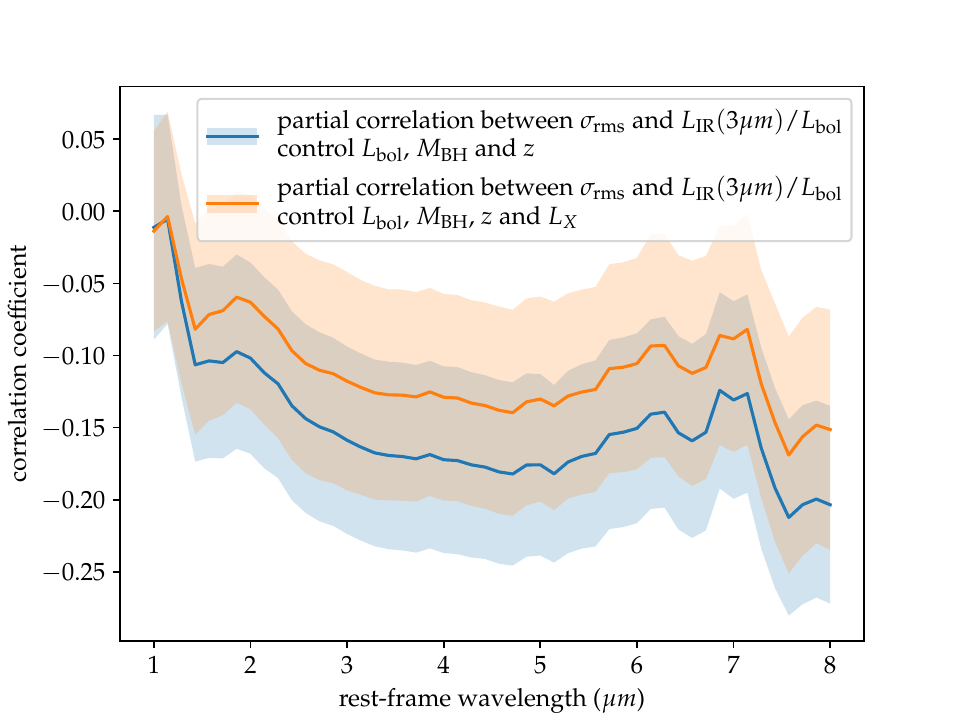}
    \caption{
        Partial correlation coefficients between $g$-band $\sigma_{\rm rms}$ and $L_{\rm IR}(\lambda)/L_{\rm bol}$ with and without controlling for X-ray luminosity.
    }
    \label{fig:xmm}
\end{figure}

Could the stronger UV/optical variability lead to the destruction of dust grains, thereby explaining the observed anti-correlation? It is known that more variable quasars tend to have stronger X-ray emission \citep{kang_var_x-ray_link_2018}. Additionally, stronger X-ray emission has the potential to destroy dust grains
through grain heating and grain charging \citep{fruchter_xray_dust_2001, gavilan_xray_dust_2022}. 
To verify this hypothesis, we matched the DR7 sample with the XMM-Newton catalogue \citep{webb_xmm_2020} using a matching radius of 5$^{\prime\prime}$ and obtain a subsample that includes 362 quasars with X-ray observations. We compute the partial correlation coefficients (controlling for the effects of redshift, bolometric luminosity and SMBH mass) between $\sigma_{\rm rms}$ and $L_{\rm IR}(\lambda)/L_{\rm bol}$ with and without controlling for X-ray luminosity for the subsample, and show the results in Fig. \ref{fig:xmm}.
Due to the much smaller size of this X-ray subsample, the resulting partial correlation coefficient has a significantly larger statistical uncertainty compared to those shown in Fig. \ref{fig:wav_corr} and \ref{fig:s82}. Nevertheless, we observe that controlling for X-ray luminosity only slightly changes the anti-correlation coefficient (from -0.16$\pm$0.06 to -0.12$\pm$0.08 at 3$\micron$, for instance). This suggests the anti-correlation between $\sigma_{\rm rms}$ and near infrared dust emission we presented in this work is not primarily due to stronger X-ray emission (which could potentially destroy dust) in more variable quasars.

On the other hand, the disk radiation is likely anisotropic because of the projection effect \citep[e.g.][]{Netzer1987}. Such effect may yield relatively weaker equatorial dust emission if the disk is viewed face-on. If the UV/optical variability, because of some reason, is also stronger when viewed face-on, the anti-correlation between variability and near infrared emission as we observed would be expected. However, in this scenario we would expect that more variable quasars have relatively weaker emission lines, which is contrary to the observations of \cite{kang_var_line_2021}.

Stronger near-infrared emission observed in less variable quasars may be more of a causal factor rather than a consequence of weaker variability, although the latter possibility cannot be completely discounted (especially if there is an underlying unknown process). How could stronger near-infrared emission suppress variability from the accretion disk? One intriguing possibility is within the framework of an inhomogeneous disk model, where a disk with weaker inhomogeneity tends to exhibit weaker variation. 
Given a certain black hole mass $M$ and bolometric luminosity $L_{bol}$,  relatively stronger infrared emission suggests that the equatorial gas reservoir feeding the supermassive black hole (i.e., the torus) is richer in dust and gas. This richness could lead to a smoother fueling of the supermassive black hole, thereby suppressing the inhomogeneity of the disk.
Further supporting this hypothesis is the observation that the anti-correlation between variability and infrared emission weakens at longer wavelengths (as illustrated in Fig. \ref{fig:wav_corr} and Fig. \ref{fig:s82}). This weakening would occur because longer infrared wavelengths experience an increase in polar dust contribution \citep{wu_ensemble_2023}. Since polar dust is situated along the polar direction and at larger scales, it shall not contribute to the accretion process, thereby mitigating the anti-correlation between variability and infrared emission.

The precise physical mechanism linking inhomogeneous feeding to variability remains unclear. One question to consider is whether the inhomogeneous feeding at larger radii has been smoothed out before reaching the inner radii. The observed correlation between the UV/optical variability of quasars and infrared emission in this study suggests that the effects of inhomogeneous feeding at the scale of the inner torus may indeed propagate to smaller radii of the disk where UV/optical emission is produced. This propagation could potentially modulate the variability observed in the UV/optical bands. Further research into the dynamics and processes occurring within the accretion disk and its surrounding environment is needed to fully understand this complex relationship.

Indeed, recent findings by \cite{Yu2020} indicate that the host galaxies of changing look AGNs, characterized by more extreme variability (see \citealt{Ren2024}) , tend to exhibit a higher fraction (albeit statistically marginal) of counter-rotating features. Similarly, \cite{Charlton2019} noted that the hosts of four changing-look quasars are predominantly disrupted or merging galaxies. These results, although based on small samples and therefore inconclusive, lend further support to the notion that the variability of AGNs could be physically linked to the feeding from the environment.

This proposed diagram, which suggests a connection between the UV/optical variability of quasars and the feeding to the accretion disk, holds promise for understanding the observed dependence of UV/optical variability on the Eddington ratio \citep[e.g.][]{wilhite_edd_ratio_variability_2008, macleod_variability_2010, Kozlowski_variability_sf_2016, Arevalo_var_edd_2023}. Here we supplement our analysis by investigating the intrinsic anti-correlation between UV/optical variability amplitude and the Eddington ratio using the samples adopted in this work. 
Previous studies typically used binning to control for the effects of other fundamental parameters, including redshift, rest frame wavelength, and SMBH mass. Instead, we attempted to use partial correlation analysis to control for the influence of other parameters.
While we observe a direct correlation coefficient of -0.33$\pm$0.02 between  $\sigma_{rms}$ and the Eddington ratio, we do find an intrinsic anti-correlation between $\sigma_{rms}$ and the Eddington ratio (with a partial correlation coefficient of -0.30$\pm$0.02). Such anti-correlation could be natural within the proposed diagram as the accretion environment of higher Eddington ratios could enable smoother fueling thus suppress the disk turbulence.
Moreover, the diagram might also elucidate why X-ray variability in AGNs shows an anti-correlation with the Eddington ratio \citep{o'neill_x-ray_var_2005,young_x-ray_var_2012,lanzuisi_x-ray_var_2014}. Furthermore, it could shed light on the observed trend where the X-ray loudness of AGNs decreases with increasing Eddington ratio \citep{kang_var_x-ray_link_2018, Duras2020, Setoguchi2024}, as disk turbulence likely regulates X-ray loudness as well \citep{kang_var_x-ray_link_2018}. Finally drawing a picture of this diagram, as \cite{wilhite_edd_ratio_variability_2008} stated, at lower Eddington ratios, ``as less gas is available, the rate at which the gas is supplied to the black hole varies more, much like the flickering of a dying fire". This analogy underscores the intricate interplay between accretion processes, environmental factors, and the observed variability in active galactic nuclei.

\section*{Acknowledgements}
We gratefully acknowledge the anonymous referee for valuable suggestions, which significantly improved the manuscript.
The work is supported by National Key R\&D Program of China No. 2023YFA1607903, National Natural Science Foundation of China (grant nos. 12033006 \& 12192221), and the Cyrus Chun Ying Tang Foundations. We thank Da-Bin Lin for helpful discussion.
 
\section*{Data Availability}

This work is based on public multi-band and time-domain photometry of SDSS quasars.


\bibliographystyle{mnras}
\bibliography{manuscript} 



\appendix

\section{Sub-samples with different redshifts}\label{AppendA}

\begin{figure}
    \centering
    \includegraphics[width=0.5\textwidth]{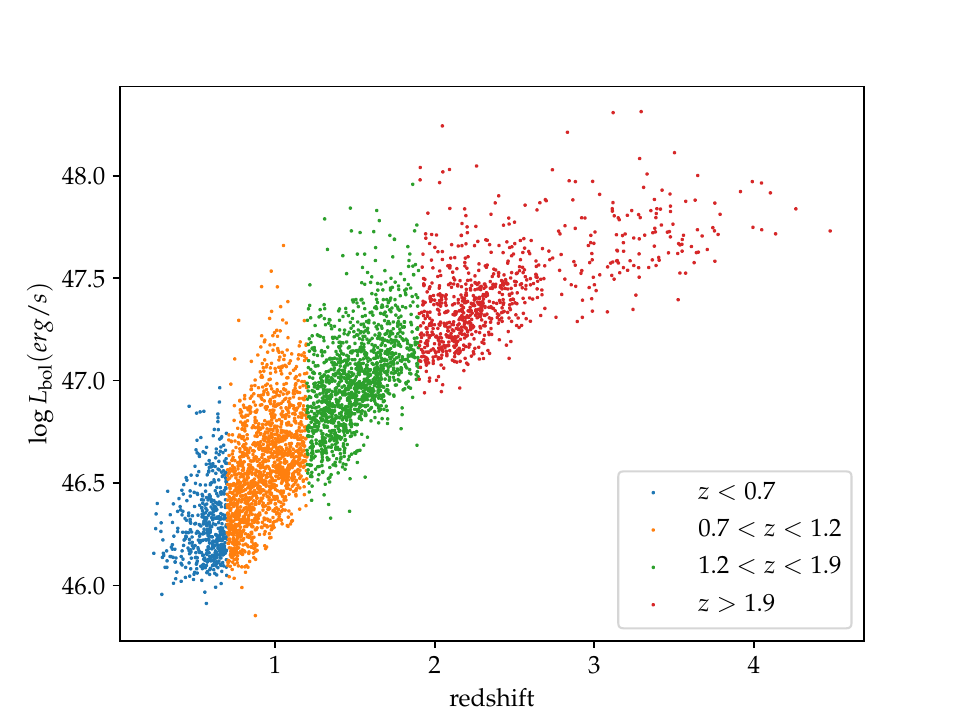}
    \caption{
        Scatter plot of full DR7 sample in the redshift - bolometric luminosity plane. Sources assigned to different redshift subsamples are colored differently.
    }
    \label{fig:redshift_select}
\end{figure}

\begin{figure*}
\centering
        
\includegraphics[width=\textwidth]{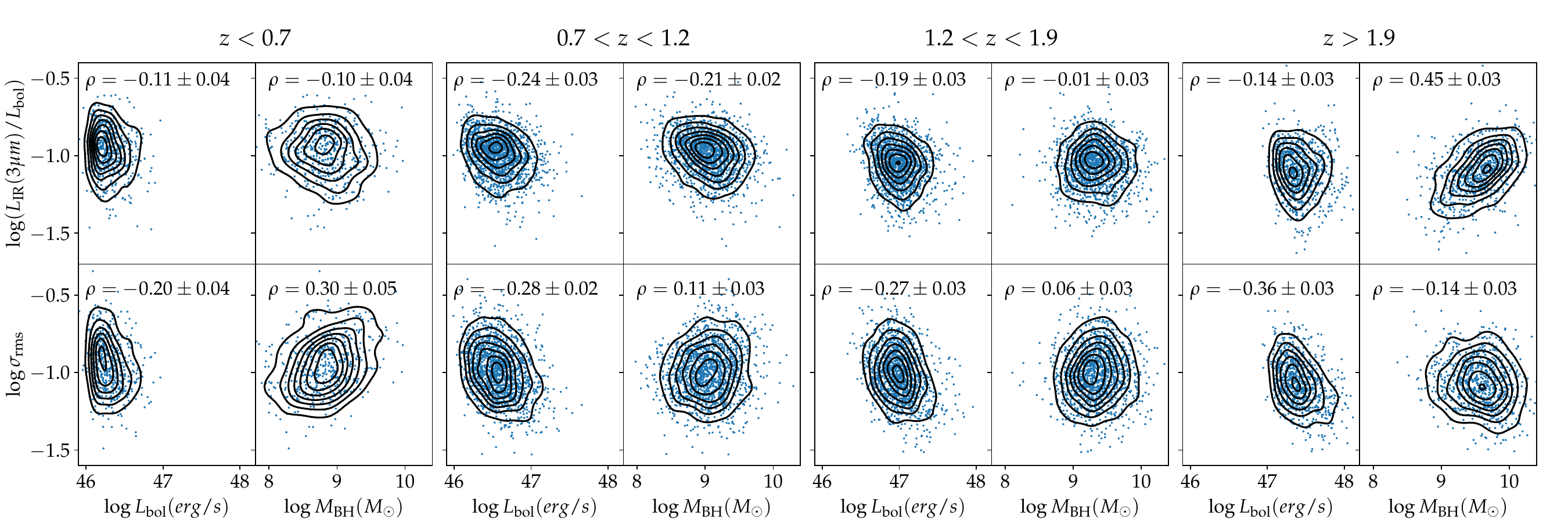} 

\caption{Same as the four left panels of Fig. \ref{fig:corrs}, but for subsamples with different redshift range.}
\label{fig:corrs_redshift}
\end{figure*}

\begin{figure*}
\centering
\includegraphics[width=\textwidth]{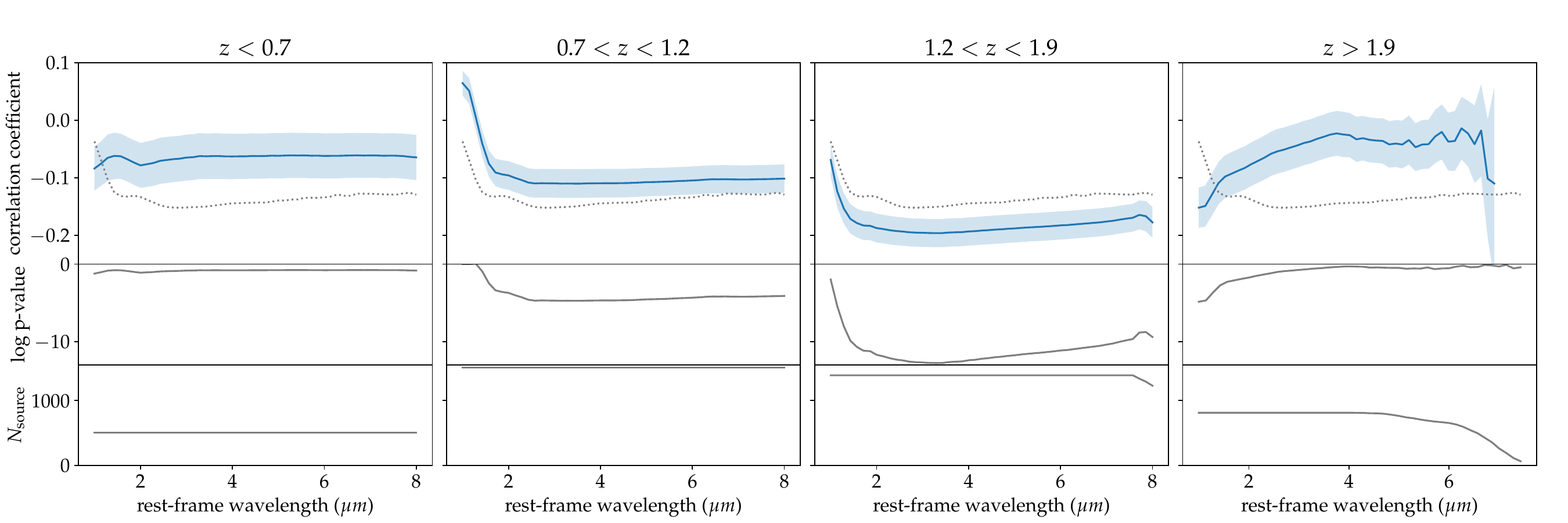}
\caption{Same as Fig. \ref{fig:wav_corr}, but for subsamples with different redshift range. Gray dotted line represent the result of full DR7 sample in Fig. \ref{fig:wav_corr}. }
\label{fig:wav_corr_redshift}
\end{figure*}

Given the large redshift range (0.18 to 4.5) of our sources, the influence of redshift on our previous conclusions cannot be overlooked. Redshift may impact our findings in several ways: the rest frame wavelength probed by a given band changes with redshift, fiducial black hole masses from \cite{shen_catalog_2011} were calculated using different spectral lines at different redshifts, and the limiting luminosity of SDSS quasars increases with redshift (see Fig. \ref{fig:redshift_select}).

To assess the impact of redshift and ensure the reliability of our conclusions, we divided the full DR7 sample (which is sufficiently large) into four subsamples based on redshift (see Fig. \ref{fig:redshift_select}): (a) 506 quasars with \( z < 0.7 \) and H$\beta$-based virial SMBH mass, (b) 1510 sources with \( 0.7 < z < 1.2 \) and MgII-based virial mass, (c) 1388 sources with \( 1.2 < z < 1.9 \) and MgII-based virial mass, and (d) 812 quasars with \( z > 1.9 \) and CIV-based virial mass. The virial SMBH masses are the fiducial masses provided by \cite{shen_catalog_2011}. The correlation analyses described in \S3 were repeated for each subsample, and the results are depicted in Fig. \ref{fig:corrs_redshift} and \ref{fig:wav_corr_redshift}.

We find some of the apparent correlations between [$L_{\rm IR}(3\micron)/L_{\rm bol}$, $g$-band $\sigma_{\rm rms}$] and [$L_{\rm bol}$, $M_{BH}$] of the subsamples within narrow redshift bins do vary compared with the full DR7 sample (Fig. \ref{fig:corrs}), and between different redshift bins. 
The mild variance in the apparent correlations is not surprising as both $L_{\rm IR}(3\micron)/L_{\rm bol}$ and  $g$-band $\sigma_{\rm rms}$ exhibit clear apparent dependence on redshift (see left panels of Fig. \ref{fig:corrs}). Additionally, $L_{\rm bol}$ of the sample is significantly redshift dependent (see fig. \ref{fig:redshift_select}), and so is the virial black hole mass $M_{BH}$.

Nevertheless, the partial correlation between $L_{\rm IR}(3\micron)/L_{\rm bol}$ and  $g$-band $\sigma_{\rm rms}$ (after controlling for $L_{\rm bol}$, $M_{BH}$ and redshift) remains statistically consistent with the full DR7 sample (see Fig. \ref{fig:wav_corr_redshift}) in subsamples a, b \& c. The only exception is subsample d ($z$ $>$ 1.9) for which the anti-correlation almost disappears at $\lambda$ $>$ 3\micron. This is likely due to the CIV-based virial black hole mass being significantly biased \citep[e.g.][]{Coatman2017} compared to the $H\beta$ or Mg II-based virial mass at lower redshifts, making it difficult to properly control for $M_{BH}$ during partial correlation analyses. Note the apparent correlation between $L_{\rm IR}(3\micron)/L_{\rm bol}$ and $M_{BH}$ (and between $g$-band $\sigma_{\rm rms}$ and $M_{BH}$) in this subsample also strikingly contradicts those from other subsamples, also suggesting the CIV-based $M_{BH}$ could be substantially biased. 

The statistically consistent patterns seen in Fig. \ref{fig:wav_corr_redshift} in redshift bins at $z$ $<$ 1.9 confirms the reliability of our key results presented in Fig. \ref{fig:wav_corr}.
Excluding the subsample with only CIV-based $M_{BH}$ from this work does not change the conclusions of this work. Interesting, excluding the subsample at $z$ $>$ 1.9 we observe a tentative evolution with redshift in the partial anti-correlation between $L_{\rm IR}(3\micron)/L_{\rm bol}$ and  $g$-band $\sigma_{\rm rms}$. This suggests the intrinsic correlation between variability and the close environment (see \S\ref{sec:discussion}) might strengthen at higher redshifts. However, since the evolution with redshift is statistically marginal (considering the error bars in the correlation coefficients in Fig. \ref{fig:wav_corr_redshift}), a robust conclusion on the potential evolution with redshift cannot yet be drawn.

\section{Sub-samples with different apparent $g$ band magnitude}\label{AppendB}

\begin{figure}
    \centering
    \includegraphics[width=0.5\textwidth]{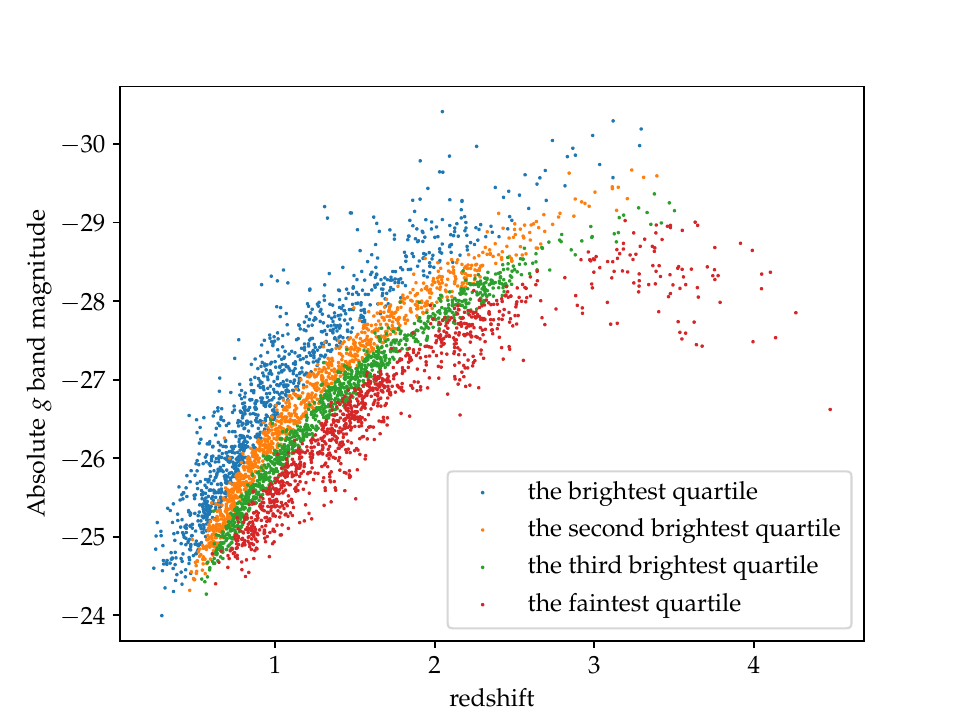}
    \caption{
        Scatter plot in the absolute $g$ band magnitude  - redshift plane, of the full DR7 sample utilized in this work. The sample is divided into four equal parts based on the observed $g$ band magnitude. 
    }
    \label{fig:g_select}
\end{figure}

\begin{figure*}
    \centering
    \includegraphics[width=\textwidth]{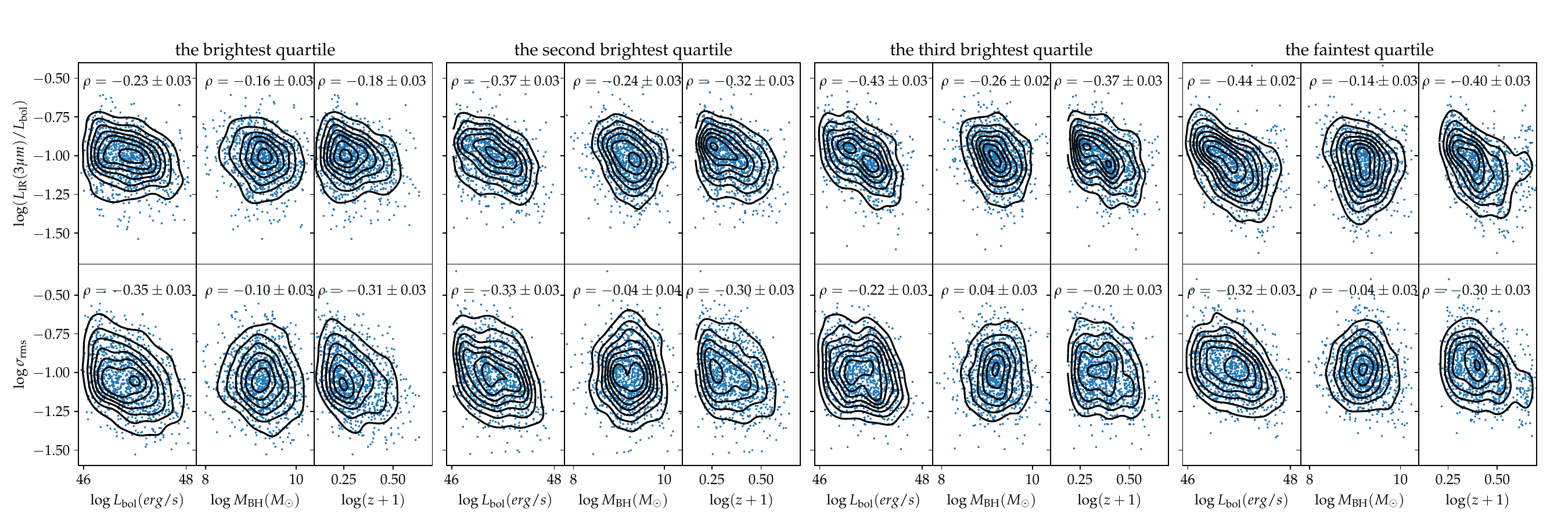}
    \caption{Same as Fig. \ref{fig:corrs}, but for the four subsamples selected based on the $g$ band magnitude (see Fig. \ref{fig:g_select}).}
    \label{fig:corrs_bf}
\end{figure*}

\begin{figure*}
\centering
\includegraphics[width=\textwidth]{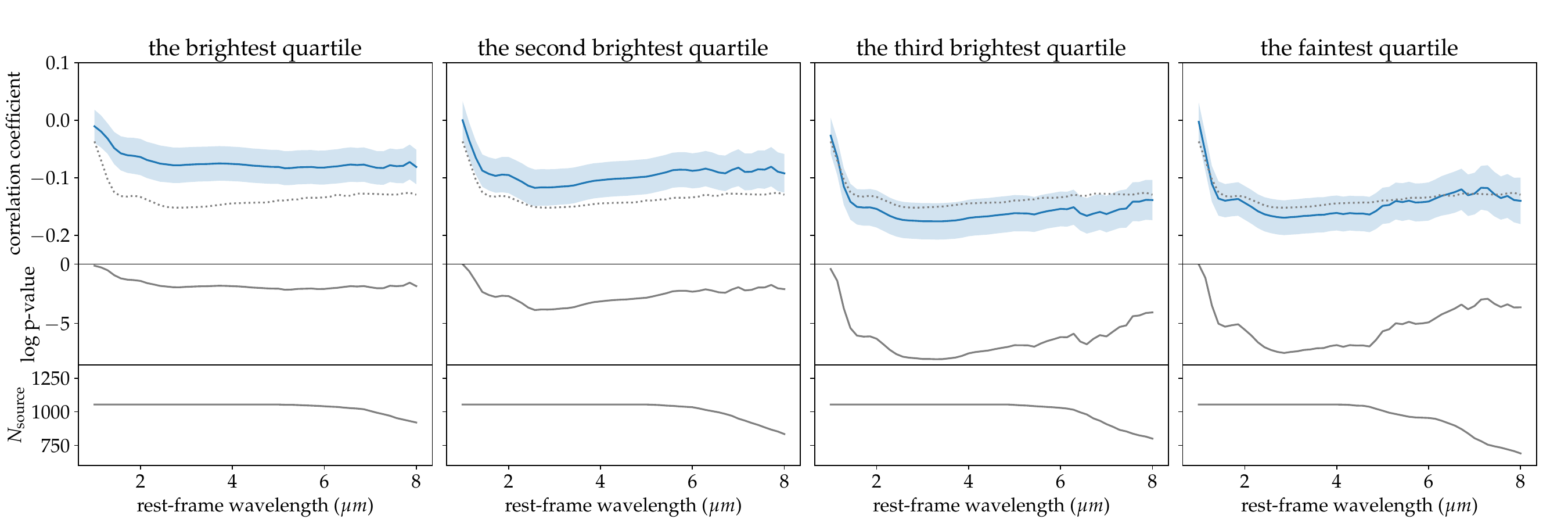}
\caption{Same as Fig. \ref{fig:wav_corr}, but for the four subsamples selected based on the $g$ band magnitude (see Fig. \ref{fig:g_select}). Gray dotted line represent the result of full DR7 sample in Fig. \ref{fig:wav_corr}. }
\label{fig:wav_corr_bf}
\end{figure*}

Photometric errors increase significantly with apparent magnitude, potentially causing the sample to behave differently at the bright and faint ends. In this section, we divided the full DR7 sample equally into four subsamples based on the sources' apparent $g$ band magnitudes, 
with each quartile contains 1054 sources. As in the previous section, we repeat the correlation analyses described in \S3 and show the results in Fig. \ref{fig:corrs_bf} and Fig. \ref{fig:wav_corr_bf}. 

As shown in Fig. \ref{fig:corrs_bf}, the apparent correlations within each subsample are generally consistent with those observed in the full DR7 sample.
In Fig. \ref{fig:wav_corr_bf}, we reproduce Fig. \ref{fig:wav_corr} for all subsamples. 
All subsamples exhibit similar anti-correlation between \( L_{\rm IR}(\lambda)/L_{\rm bol} \) and excess variance \( \sigma_{\rm rms} \), indicating our results are not significantly affected by larger photometric errors for fainter sources.  
Meanwhile, we find marginally weaker anti-correlation in the brightest quartile and the second brightest quartile. This observation may be attributed to the lower redshifts of brighter sources (see Fig. \ref{fig:g_select}), as Fig. \ref{fig:wav_corr_redshift} illustrates that low-redshift sources may tentatively exhibit weaker anti-correlation.


\bsp	
\label{lastpage}
\end{document}